\documentclass[aps,prb,showpacs,twocolumn,superscriptaddress]{revtex4-1}
\usepackage{comment}
\usepackage{amsmath}
\usepackage{amsfonts}
\usepackage{amssymb}
\usepackage{graphicx}
\usepackage{epstopdf}
\usepackage{epsfig}
\usepackage{subfigure}
\usepackage{natbib}
\usepackage{hyperref}
\usepackage{float}

\begin{document}
	
\title{The limits of rechargeable spin battery}

\author{A. V. Yanovsky}
\thanks{andrey.v.yanovsky@gmail.com}
\affiliation{B. Verkin Institute for Low Temperature Physics and Engineering of the National Academy of Sciences of Ukraine, 47 Nauky Ave., Kharkov 61103, Ukraine}

\author{P. V. Pyshkin}
\thanks{pavel.pyshkin@gmail.com}
\affiliation{Department of Physical Chemistry, University of the Basque Country UPV/EHU, 48080 Bilbao, Spain}
	
	\begin{abstract}
		 We discuss how the ideal rechargeable energy accumulator can be made, and what are the limits for solid state energy storage.
		 We show that in theory the spin batteries based on heavy fermions can surpass the chemical ones by energy capacitance.
		 The absence of chemical reactions in spin batteries makes them more stable, also they don't need to be heated in cold conditions.
		 We study how carriers statistics and density of states affect energy capacity of the battery.
		 Also, we discuss hypothetical spin batteries based on neutron stars.   
	\end{abstract}
\maketitle
\section{Introduction}

Rechargeable electric batteries are one of the most important devices of modern civilization. It is obvious that their role will only increase in the future. Unfortunately, the existing chemical rechargeable batteries (based on reversible electrochemical reactions) are far from ideal. This manifests itself, for example, in their inevitable irreversible degradation, slow charging, relatively low energy capacitance per unit mass, the need for heating when the temperature drops, etc.
Of course, the progress doesn't stop, however the most efforts now are applied to chemistry and physical chemistry properties of batteries (see \cite{Chen:2020, Suga:2011}).
Relatively new physical idea of quantum battery explores quantum state and entanglement properties \cite{Alicki:2013,Binder:2015,Fe-Co:18,Barra:2019}.
The idea of using spin degree of freedom to store energy attracts a lot of attention last years~\cite{Tian2017,Xie2018,Bozkurt2018,Nguyen2020}.
Particularly, in recent article~\cite{battery} authors propose a spin battery (SB) which is half-metal spin valve with suppressed spin flips of conducting electrons. This solution would allow us to store reversibly the electric energy without any chemical reactions at the charging process using nonequilibrium states of quasiparticles in a conductor instead.

Hence, the following question appears: is it possible for SB to surpass chemical battery, and what are the properties of ideal SB? In this paper we theoretically ``test'' possible limits of solid state for SB-s and more exotic matter of neutron stars as well.

\section{Chemical potential}

\begin{figure}[t]
	\centering
	\includegraphics[width=0.7\linewidth]{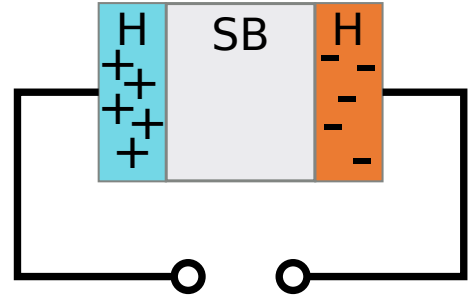}
	\caption{Schematic illustration of spin battery (denoted by ``SB'' in picture) which is a conductor between half-metal electrodes (denoted by ``H'' in picture) with opposite spin polarization. 
	}
	\label{fig-scheme}
\end{figure}

The energy in SB is stored in spin carriers' density deviation from its equilibrium value under condition that spin relaxation is suppressed in this conductor.
In other words, such a battery is just a certain volume filled with spin particles, and energy accumulation appears due to a development of the non-equilibrium spin state. 
Such spin particles can be electrons in conductor~\cite{battery}, whereas being charge carriers, and their spin direction~$\pm$ can be determined by the external magnetic field or magnetic contacts.
In order to charge such a battery containing charged spin carriers or to transfer accumulated non-equilibrium spin concentration into charge current one can use for instance antiparallel magnetized half-metal~\cite{Picket:2001,Mazin:2000,CongChen:2019} electrodes which passes only~$+$ or~$-$ spin correspondingly~\cite{battery}, see Fig.~\ref{fig-scheme}.

In this situation the charging potential difference~$\delta \varphi$ induces variations of chemical potentials of $\pm$~components (after the charging time when the equilibrium established):
$$\mu_{\pm} = \mu_{0,\pm} + \eta_{\pm} \ ,$$
where $\mu_{0,\pm}$ are equilibrium electrochemical potentials of the discharged battery determined by density of correspondent $\pm$~components.  
Chemical potentials $\eta_{\pm}$ are induced by charging process, and their values could be found from the conditions $\delta \varphi = \eta_+/q_+ - \eta_-/q_-$ (where $q_\pm$ are electric charges of corresponded carriers), together with spatial electroneutrality which follows from Poisson equation $\Delta \varphi = - 4 \pi \{q_+ [\rho_+(\mu_+) - \rho_+(\mu_0)] + q_- [\rho_-(\mu_-) - \rho_-(\mu_0)]\}$ (with RHS equal to zero):
\begin{equation}
q_+ [\rho_+(\mu_+) - \rho_+(\mu_0)] + q_- [\rho_-(\mu_-) - \rho_-(\mu_0)] = 0
\label{eq-neutrality-gen}
\end{equation}
where $\varphi$ is electrical potential, $\rho_\pm$ are the densities, $q_\pm$ are charges of correspondent spin~$\pm$ carriers. Here, for simplicity, we assume $\mu_{0,+} = \mu_{0,-} = \mu_0$, as it is obvious that the equilibrium value of electrochemical potential does not affect the basic principles of energy storage.

\begin{figure}[t]
	\centering
	\includegraphics[width=0.9\linewidth]{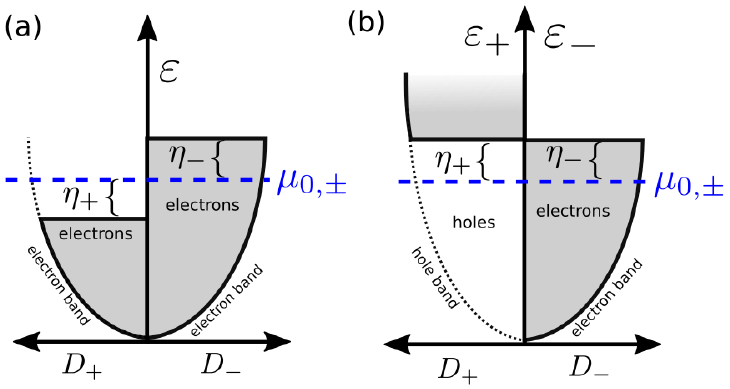}
	\caption{(a) One-band ``purely-electronic'' spin battery. (b) Two-band ``electron-holes'' spin battery. Filled levels are drawn in gray color. The number of carriers of one spin decreases when the number of carriers of another spin increases for one-band battery, and polarity is not important. We chose polarity of two-band battery in such a way to have increasing of both types of carriers during charging process.
	}
	\label{fig-filling}
\end{figure}

For one-band conductor we have $q_+ = q_- = e$, where $e$ is the electron charge. 
Moreover, we can consider the usage of two-band conductors which contain not only the electrons but also ``holes'' with opposite charge. What is more, the carriers are polarized in such a way as to spin~$\pm$ connected to charge~$\pm$. 
In this situation the charges $q_\pm$ have the opposite sign: $q_+ = -q_- = e$. This is possible when interaction between carriers from different bands is weak~\cite{ZAYETS201853}, or by using electron-hole pairing methods~\cite{Shevchenko:1976, PhysRevLett.93.266805, Nandi:2012} for the spin-flip suppressing mentioned in~\cite{battery}. When the one-band battery is being charged it follows to increasing the number of certain spin carriers, and it necessarily leads to decreasing the number of opposite spin carriers in order to preserve electroneutrality~(\ref{eq-neutrality-gen}).
In two-band battery we always chose polarity of charging voltage in such a way to have increasing of the number of carriers of both spins ($\eta_+ > 0$). It means that we connect positive-guided contact to the electrode with ``$+$''-polarity (corresponding to ``holes'' with charge $q_+$), and negative-guided contact to the electrode with ``$-$''-polarity (corresponding to electrons with charge~$q_-$).
Non-equilibrium states caused by deviations of spin densities in one-band and two-band SB-s are shown in Fig.~\ref{fig-filling}. Here we show schematically the filling energy levels $\varepsilon$ as functions of corresponded densities of states (DOS) $D_\pm$ for spins~$\pm$. 

As can be seen, two-band battery has polarity, and such SB is equal to chemical battery but with the following difference. In chemical battery we have the concentrations changing, and correspondingly the changing of chemical potentials with respect to electrodes, while $q_\pm$ corresponds to the one ion charge: $-q_+ = q_- = -z F/N_a$, where $z$ is positive valency, $F$ is Faraday constant, and $N_a$ is Avogadro number (for definiteness we chose the sign the same as elementary charge has, we assume the same absolute value of valency of all ions).
In the absence of charging potential difference in the circuit, the presence of non-equilibrium chemical potentials leads to appearance of diffusion forces, which pull-in or push-out charges into the circuit. It happens on electrodes of the opposite ``affinity'' (such an affinity is related to chemical reactions in usual chemical battery, or it is related to the presence of conduction band only for certain spin on $\pm$ electrodes in the SB).
Asymmetry of charge moving during relaxation into thermodynamic equilibrium state causes electric current in the full circuit~\cite{battery}.
Also, we can consider SB with spin/charge carriers are not being usual conductive electrons but quasi-particles. Such quasi-particles even can have zero electric charge, but in this situation the movement of such quasi-particles do not cause electric current, and thus the energy extraction from this battery is difficult. 
SB doesn't require chemical reactions, and therefore it does not suffer from chemical degradation. 
As we show below, SB doesn't require heating in the case it consists of degenerate gas of charge/spin carriers.
Obviously, SB can be a source not only of charge but also of spin current~\cite{Awschalom:2001,Brataas:2002,Long:2003}\footnote{Attaching usual conductor to a conductor with non-equilibrium spin distribution will cause diffusive spin current which equalizes spin concentrations.}.  
At last, SB can be charged without electrodes with using polarized electromagnetic radiation~\cite{polarizedlight}.

\section{General formulas for the energy of charged SB}

Let us denote $E_\pm(\mu_\pm)$ as the total internal energy of carriers of~$\pm$ components for given value of electrochemical potential $\mu_\pm$. The energy stored in the battery is a difference between the internal energies of charged and discharged states 
\begin{equation}
\delta E = E_+(\mu_{0} + \eta_+) + E_-(\mu_{0}+\eta_-) - E_+(\mu_{0}) - E_-(\mu_{0}).  
\label{eq-E}
\end{equation}
At microscopic level the value of~$E$ in SB as well in chemical battery is determined by equilibrium energy distribution of carriers~$n_{\pm}(\varepsilon, \mu)$, DOS~$D_\pm(\varepsilon)$ and by the volume~$\Omega$:    
\begin{equation}
E_\pm(\mu_\pm) = \Omega \int\limits_{0}^{\infty} d \varepsilon  \varepsilon  D_\pm(\varepsilon)  n_\pm(\varepsilon, \mu_\pm) \ .
\label{eq-energy1}
\end{equation}
Also, we can write the following expression for $\rho_\pm$ in order to substitute it in~(\ref{eq-neutrality-gen}) 
\begin{equation}
\rho_\pm(\mu_\pm) = \int\limits_{0}^{\infty} d \varepsilon D_\pm(\varepsilon)  n_\pm(\varepsilon, \mu_{\pm}) \ .
\label{eq-dencity1}
\end{equation}
As can be seen from (\ref{eq-E}) and (\ref{eq-energy1}), the energy distribution determines which parts of the DOS dependencies~$D_\pm(\varepsilon)$ give the main contribution.  
Of course, the energy distribution can be classical (Boltzmann), Bose, Fermi or, even, so-called fractional statistics.

Note, usually, $\varepsilon D(\varepsilon)$ increases with $\varepsilon$, and on the other hand Bose distributions collect particles in states with lower energies where DOS has its minimal values. 
Of course, DOS of Bose distributions can have certain singularities (see \cite{Pastur:1982}), thus some low dimensional systems require special investigation.
However, in general case, if we do not consider non-physical DOS which diverge at zero faster than $\varepsilon^{-1}$, then in order to have maximal capacitance of SB, particles should occupy more states with larger DOS.
In this regard, fermions are the best choice for SB-s. As two fermions cannot occupy the same state, they have to fill levels with higher energy with the number of particles on the rise. 
Below we discuss the influence of quantum statistics in details.
Intuitively we understand that battery is more sensitive to the temperature in the case of classical statistics. 
In general case, when $\mu_0$ and $q_\pm$ are given, the searching of a battery with maximal energy capacitance is reduced to variational problem -- the searching for conditional maximum of the functional~$\delta E [D_\pm, n_\pm]$.
This functional is linear on $D_\pm$ and $n_\pm$, while restrictive condition is the equation (\ref{eq-neutrality-gen}).
In reality distributions and density of states are limited by physical states mentioned above.

\section{Briefly on classical statistics}

Boltzmann statistics corresponds to sufficiently low densities $\rho_\pm$, when $\exp\{-\mu/T\} \gg 1$, where $T$ is the temperature in energy units.
Note, in the case $q_+ = q_-$ the electroneutrality condition (\ref{eq-neutrality-gen}) is broken when $|e\delta \varphi| \sim T$.
In this situation increasing of $\pm$ component (depending on the sign of $\delta \varphi$) can not be compensated via decreasing of $\mp$  component due to the insufficient amount of it, and thus the equation (\ref{eq-neutrality-gen}) does not have a solution.
Since the temperature $300 \text{K}$ approximately corresponds to the potential difference only $0.03 \text{V}$, the case $q_+ = q_-$ is out of interest in classical limit.
Particularly, this is why there is no sense in using of chemical accumulators with different types of cations or anions in such a way that the opposite electrodes make unbalance of different ions of the same sign during the charging, because electroneutrality breaks at relatively low voltages. 
Generally, electroneutrality is not broken when $q_+ = - q_-$ with increasing of $\delta \varphi$. In such a case only the average value of DOS matters.    
Really, the classical distribution $n_\pm(\varepsilon, \mu) \approx \exp\{\mu/k T\} \exp\{-\varepsilon/ k T\} $, where constant in $\mu$ is chosen to satisfy normalization condition for the density of particles.  
From (\ref{eq-neutrality-gen})--(\ref{eq-dencity1}) we have the following
\begin{equation}
\delta E \approx \Omega \Big[\exp\{\frac{\eta_+}{T}\} - 1\Big] \Big(E_+(\mu_{0}) + E_{-}(\mu_0)\Big).
\label{eq-energy-class}
\end{equation}
Accordingly, the value of $\delta E$ in the leading approximation with respect to $T|q\delta \varphi|^{-1} \ll 1$ is  
$$\delta E \sim \Omega T [\rho_0(T, \mu_{0} + \frac{q\delta\varphi}{2}) - \rho_0(T, \mu_0)] \ ,$$
where $\rho_0(T, \mu_0)$ is the equilibrium concentration for given temperature. 
As we can see, singularities of $D(\varepsilon)$ don't play role in the case of classic statistics, and only the average of DOS appears in~(\ref{eq-energy-class}). 
Also, one can see that $\delta E$~strongly depends on~$T$, and on concentrations as well which reduces application of such accumulators at low temperatures without additional heating. 
Let us show that Fermi statistics changes the situation dramatically.  

\section{``Fermi'' batteries}

Quantum Fermi statistics can be realized in SB-s made from degenerate conductors like metals where equilibrium electrochemical potential approximately equal to Fermi energy $\mu_0 \simeq \varepsilon_F$. 
Usually in metals Fermi energy is of order $10\text{eV} \sim 10^5\text{K}$.
Therefore we can consider energies $|q_\pm \delta \varphi| \sim |\eta_\pm| < \mu_0$ and neglect thermal blurring.
The energy of non equilibrium state with splitting of spin components and corresponded Lagrange function for this approximation can be found in \cite{Pyshkin:2014}. Here we write it with using our notation:
\begin{equation}
\delta E \approx \Omega \Big[\int_{\mu_{0}}^{\mu_{0} + \eta_+} d \varepsilon \varepsilon D_+(\varepsilon)  + \int_{\mu_{0}}^{\mu_{0} + \eta_-} d \varepsilon \varepsilon D_-(\varepsilon)\Big] \ , \label{eq-energy-fermi}
\end{equation}
\begin{multline}
q_+ [\rho_+(\mu_+) - \rho_+(\mu_0)] + q_- [\rho_-(\mu_-) - \rho_-(\mu_0)] \cong \\q_+ \int_{\mu_0}^{\mu_0 + \eta_+} d \varepsilon D_+(\varepsilon) + q_- \int_{\mu_0}^{\mu_0 + \eta_-} d\varepsilon D_-(\varepsilon) = 0\ .
\label{eq-neutrality-fermi}
\end{multline}
As can be seen from (\ref{eq-energy-fermi}) and (\ref{eq-neutrality-fermi}) the Fermi level shift determines the limits of integration, and as we show below the DOS singularities can play an important role in the properties of ``Fermi'' battery.
Note, in one-band battery ($q_+ = q_-$ ) integrand in (\ref{eq-energy-fermi}) coincides up to multiplier $\varepsilon$ with integrand in RHS of (\ref{eq-neutrality-fermi}), which is set to zero.  
Thus, in this case we have certain compensation when we make small parameter expansion for $\eta_{\pm}/\mu_{0}$.
Physical meaning of that is the following: for one-band battery we lose in energy when decrease the number of quasiparticles of a certain kind during charging, for two-band battery the number of quasiparticles of both kind only increases during charging (when polarity is correctly chosen).

Firstly, we consider the case when DOS does not have any singularities, i.e. $D_\pm(\varepsilon)$ are analytical functions which can be expanded in Taylor series near initial Fermi level $\mu_0$: $$D_\pm(\varepsilon) \approx D_\pm(\mu_0) + (\varepsilon - \mu_0) D'_\pm(\mu_0) + \frac{(\varepsilon - \mu_0)^2}{2} D''_\pm(\mu_0) + \dots \,$$ 
where prime means differentiation according the only argument.

In the case of smooth DOS, SB based on one-band and two-band normal metal have quite different energy storage properties.   
In one-band metal we have $D_+(\mu_0) = D_-(\mu_0) = D(\mu_0)$, and the same relations for derivatives $D', D''$. 
The electro-neutrality condition for the case $q_+ = q_-$ with second-order accurate respecting to $\eta_{\pm}/\mu_0$ gives the following equation:
\begin{equation}
\eta_+ + \eta_- + \frac{D'(\mu_0)}{2 D(\mu_0)} (\eta_+^2 + \eta_-^2) = 0
\label{eq-eta-expanded},
\end{equation} 
which leads to
\begin{equation}
\eta_- \approx - \eta_+ - \frac{ D'(\mu_0)}{ D(\mu_0)} \eta_+^2  + \dots.
\end{equation}
Thus, from (\ref{eq-energy-fermi}), (\ref{eq-neutrality-fermi}), (\ref{eq-eta-expanded}) the energy of one-band SB is
\begin{equation}
\delta E \approx \Omega D(\mu_0) \eta_+^2 + \dots.
\label{eq-energy-smooth-dos}
\end{equation}
Here dots mean high-order terms with respect to $\eta_{\pm}/\mu_0$.
As can be seen from (\ref{eq-energy-smooth-dos}), the linear terms (with respect to $\eta_\pm$) are canceled, and the energy has only quadratic dependence on potential difference. This fact leads to the appearance of small multiplier $\sim e \delta\varphi/\mu_0$ compared with two-band SB.
Indeed, in two-band SB the linear approximation with respect to $\eta_{\pm}/\mu_0$ is enough, which gives $\eta_+ = \eta_-$, and we have
\begin{equation}
\delta E = 2 \Omega D(\mu_0) \mu_0 \eta_+ \ , \ \ \frac{\delta E_{1-band}}{\delta E_{2-band}}\sim \frac{\eta_+}{\mu_{0}}.
\label{eq-energy-2band}
\end{equation}
As can be seen from (\ref{eq-energy-2band}) the energy is linear and greater than in one-band case.
In both cases the higher Fermi energy (i.e. the density of carriers), the higher energy of SB.  
In one-band SB  the energy is proportional to DOS at $\mu_0$, while in two-band SB the energy is proportional to product of DOS at $\mu_0$ and $\mu_0$ itself. As can be seen, in both cases (\ref{eq-energy-smooth-dos}) and (\ref{eq-energy-2band}) we have DOS at Fermi level without averaging over energy  as opposed to classical case.

\begin{figure}[t]
	\centering
	\includegraphics[width=0.6\linewidth]{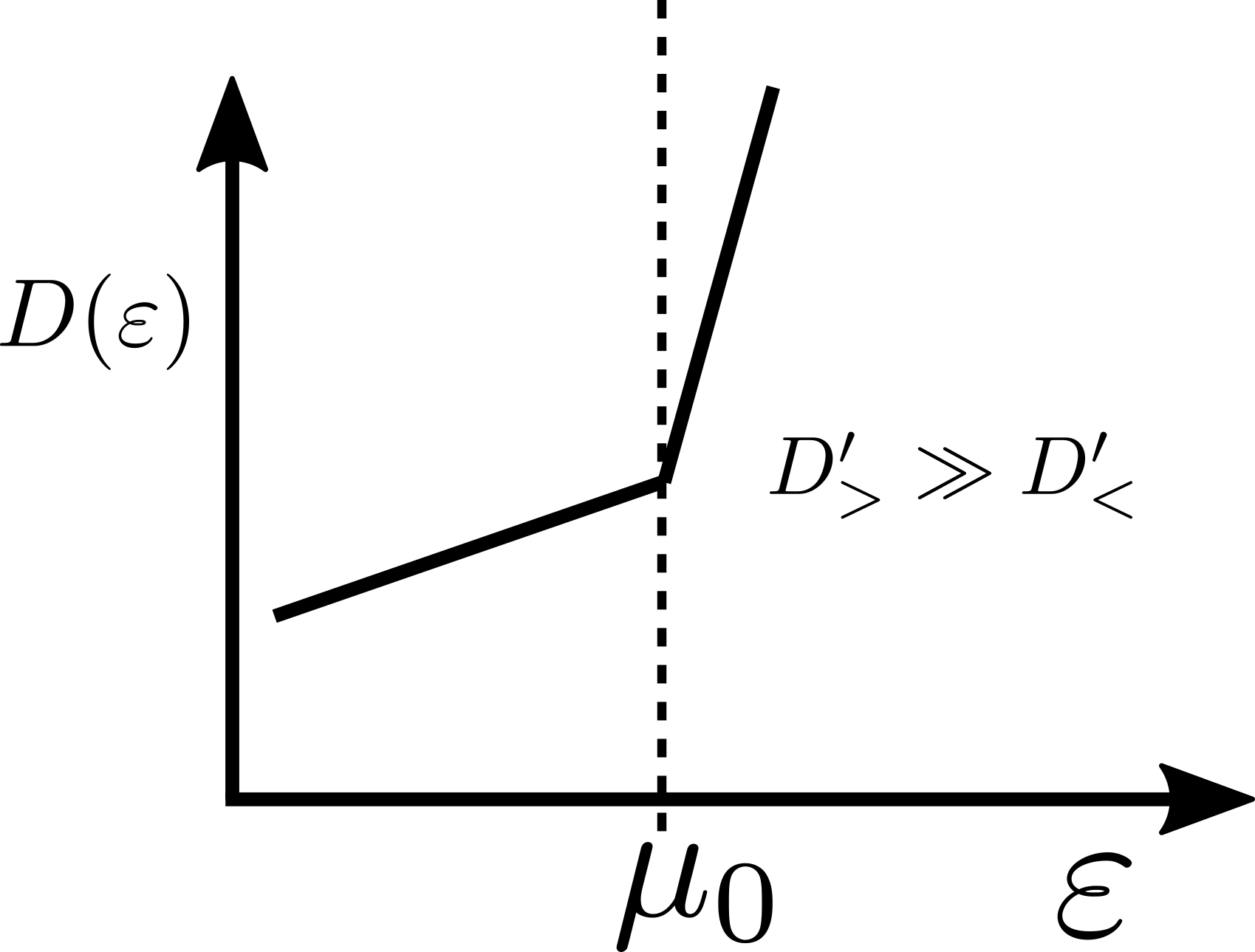}
	\caption{
	Model DOS cusp at Fermi level with high value of DOS derivative at right.}
	\label{fig-singularity}
\end{figure}

Let us discuss how DOS singularities can increase power capacity of one-band Fermi SB based on discussed above high-value parameter which contains DOS derivative.
Really, it is not unusual to have cusps in $D(\varepsilon)$, for instance, when Van Hove singularities are present~\cite{VanHove,Bassani} etc.
For the sake of simplicity we consider the model case of linear dependence the right and left of the cusp, and also we assume the cusp of $D(\varepsilon)$ is exactly aligned at $\mu_0$. 
\begin{multline}\label{cusp-1}
D(\varepsilon) = D(\mu_0) + D'(\mu_0 + 0)(\varepsilon - \mu_0) \theta(\varepsilon - \mu_0) + \\ D'(\mu_0 - 0)(\varepsilon - \mu_0) \theta(\mu_{0} - \varepsilon)  \ .
\end{multline}
Note, the expression~(\ref{cusp-1}) is not an expansion with respect to the small parameter, but the model DOS. Let us denote $D = D(\mu_0)$, $D_{>}' = D'(\mu_0 + 0)$ and $D_{<}' = D'(\mu_0 - 0)$, see Fig.~\ref{fig-singularity}.  

After taking integrals in equations (\ref{eq-neutrality-fermi}), (\ref{eq-energy-fermi}) we obtain
\begin{equation}
D \eta_+ + D'_> \frac{\eta_+^2}{2} + D \eta_- + D'_< \frac{\eta_-^2}{2} = 0
\end{equation}
\begin{equation}
\delta E = \left[D \frac{\eta_+^2}{2} + D'_> \frac{\eta_+^3}{3} + D \frac{\eta_-^2}{2} + D'_< \frac{\eta_-^3}{3}\right]\Omega. 
\end{equation}
Assume the value of $D_>'$ is big enough to have $D_>'\eta_+/D \gg 1$, while derivative $D'_<$ is small (for simplicity we can take it zero).
In such a case $\eta_- \approx - 2^{-1}D^{-1} D_>'\eta_+^2$, and the energy $\delta E \sim \Omega D'_>\eta_+^3$.
Using the condition $e \delta \varphi = \eta_+ - \eta_-$ we can estimate the derivative value of $D'_>$ which corresponds the situation when the energy of a Fermi battery with cusp exceeds estimation~(\ref{eq-energy-smooth-dos}) for the same value of carriers density and the same DOS at Fermi level.   
Let us take $\varphi \sim 1 \text{V}$, $\mu_0 \sim \varepsilon_F \sim 10 \text{eV}$, and then
$$ D'_>\eta_+^3 > D \eta_+^2 \Rightarrow \frac{D'_{>} \eta_+}{D} > 1 \Rightarrow \frac{D}{D'_{>}} < 1 \ \text{eV,}$$
which looks experimentally feasible.
The possibility of such power capacity increasing follows from that fact that DOS derivative gives the {\em main contribution} to the energy instead of being corrections.   

\section{Heavy fermions}

The density of states of conduction electrons in metals are high as a result of high density~$\rho$.
Obviously, DOS in normal metals is greater than DOS in liquid electrolytes, despite small effective electron mass in metals.
In fact, the smallness of effective electron mass is compensated by high Fermi velocity, which in several orders greater than, for example, thermal ions velocities in dilutions.

One can imagine even more effective conductors for SB-s, which contain so called ``heavy fermions''~\cite{Alekseevskii,Moshchalkov,Stewart,Ott,Heavy-ferm-1-Feng2021,Heavy-ferm-2-Chatterjee2021}, like some intermetallic antiferromagnetic alloys with f-electrons.
Indeed, in such correlated conductors the effective masses $m_{heavy}^*$ of carriers are $100 \div 1000$ times greater than effective masses in normal metals $m^*_{normal} \sim 1 \div 10 \ m$, where $m$ is a free electron mass.

\begin{figure}[t]
	\centering
	\includegraphics[width=0.6\linewidth]{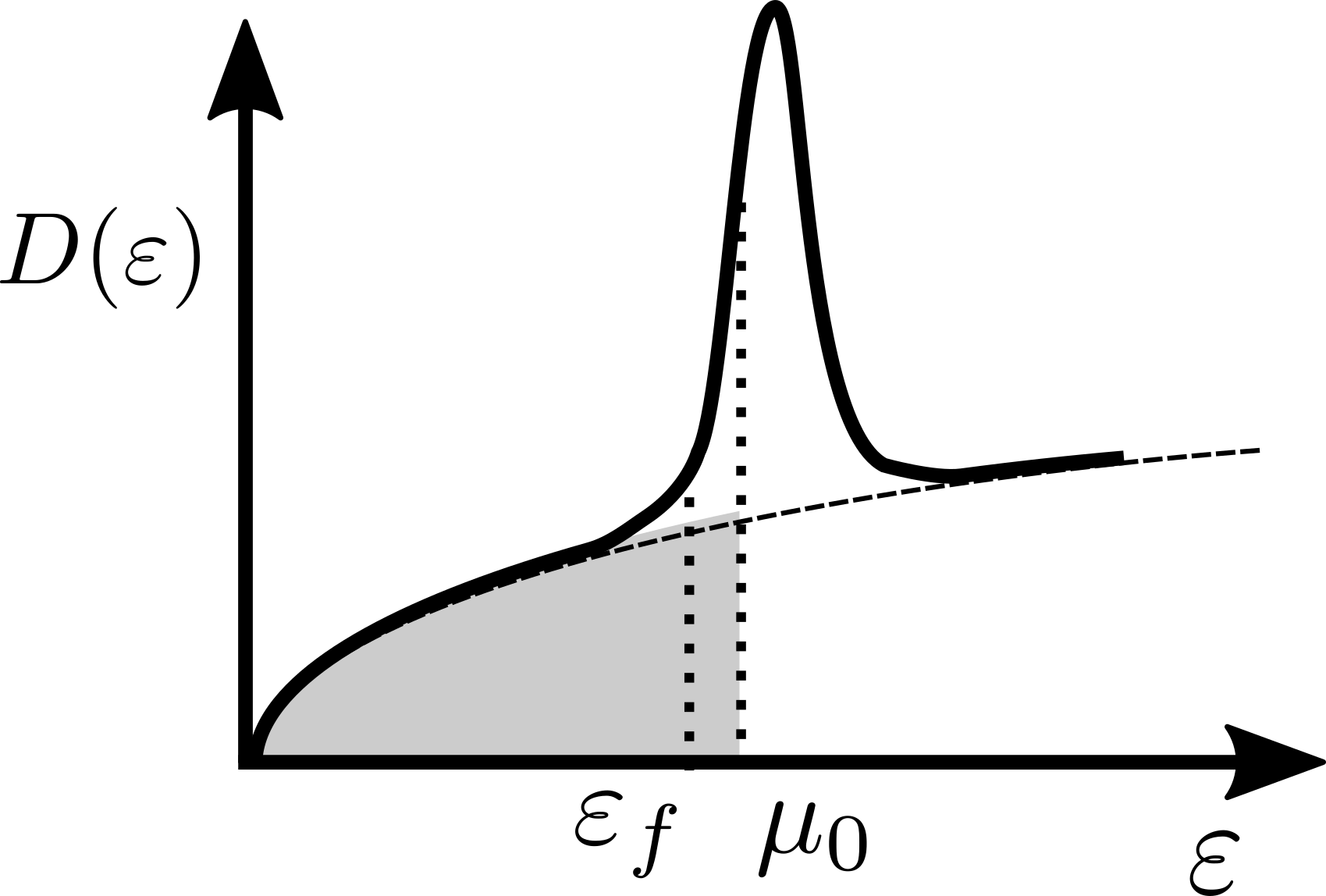}
	\caption{
	Schematic illustration of DOS in intermetalic alloys with f-electrons. Dotted curve corresponds to $D(\varepsilon)$ at absence of sf-interaction. ``Heavy fermions'' form states near the peak. $\varepsilon_f$ is the electron bound energy in f-shell.}
	\label{fig-heavy}
\end{figure}

The DOS of heavy fermions has irregularity, see Fig.~\ref{fig-heavy}.
If we substitute heavy fermion DOS in expressions (\ref{eq-energy-smooth-dos}) and (\ref{eq-energy-2band}) taking into account that $\eta_\pm \sim e V$, $\mu_0 \sim \varepsilon_F$, $D_{heavy} \sim (m^{*}_{heavy}/m^{*}_{normal})^{3/2}_{heavy} D_{normal} \sim 10^{4}  D_{normal}$ we get the following rough estimation
\begin{equation}
\delta E_{heavy} \sim 10^4 \cdot \frac{\Omega}{\lambda_F^3} \frac{e V}{\varepsilon_F} \begin{cases}
eV \ , \ \ \text{one-band battery} \\ \varepsilon_F  \ , \ \ \text{two-band battery}
\end{cases}
\end{equation}
Given that in metals we have $\lambda_F \sim 10^{-10}$ meters, and Fermi energy is being in range from one to ten electron-volts then for $V \sim 1$ Volt per $1 \text{cm}^3$ we estimate maximal power capacity of such SB: $W_{max} = 3600^{-1} \delta E_{max}/V \sim 10^4 \text{Ah}$.
For comparison, the battery power capacity of Apple iPhone 12 Max Pro indicated on specifications is $3.7\text{Ah}$, i.e. three orders of magnitude less than one cubic centimeter of hypothetical heavy fermion SB.  
Given that metal density is approximately $10^4 \text{kg}/\text{m}^{-3}$ we obtain the energy density of heavy fermion metallic SB is about $100 \div 1000$ Megajoules per kilogram. This value can be one order of magnitude greater than normalized energy density of gasoline as a fuel for internal combustion engine (which in its turn is about $46$ Megajoules per kilogram \cite{GOST, GOST2}).
At the same time the energy density of chemical batteries is two orders of magnitude worse than the gasoline, and this fact is an important problem of modern electric transport. 

\section{Neutron star as a record SB}

Since neutrons in neutron stars are compressed by gravitational force to huge density, and due to the fact they are fermions, they gain huge DOS values. 
Neutron has spin but does not have a charge, and starting from this fact we can imagine supercivilization which could store energy by ``charging'' neutron stars via polarized radiation, i.e. producing the difference of concentrations of spins referred to certain direction in space. 
Here we don't discuss the problem of extraction of this energy. It is impossible to transform this energy into electric current due to the electrical neutrality of neurtons. Also, it is obvious that it is impossible to make some contacts with neutron star. 
Thus, the supercivilization has to investigate non-contact method. Nevertheless, let us here estimate colossal energy which can be stored in neutron star in this fantastic scenario. 

If neutron star has a Sun mass $M_\odot \approx 2\cdot 10^{30}\text{kg}$, its radius should be $R \approx 10\text{km} = 10^4\text{m}$, and correspondingly the density $\rho = 3 M_\odot/4 \pi R^3 \approx 1.4 \cdot 10^{18}\text{kg}/\text{m}^{-3}$.
These numbers are taken just for understanding the scale of such a fantastic ``device'', the mass of a real neutron star should be greater than Tolman–Oppenheimer–Volkoff limit~$2.17 M_\odot$~\cite{Oppenheimer,Tolman}.  
In order to estimate DOS and Fermi energy one can use simple Thomas-Fermi formulas with relativistic corrections~\cite{neutronstars1, neutronstars2} because Fermi velocity is close to speed of light~$c$ for such a huge densities, and moreover, the momentum  $p \gg m_n c$, where $m_n$ is a neutron mass. Thus we have
\begin{eqnarray}
\varepsilon &=& c \sqrt{p^2 + m_n^2 c^2} - m_n c^2 \approx c p\ , \ \ p \gg m_n c \,  \\
D(\varepsilon) &=& \frac{\varepsilon^2}{\pi^2 c^3 \hbar^3}, \, \\
\varepsilon_F &\approx& \pi c \hbar \Big(\frac{3 \rho}{\pi m_n}\Big)^\frac{1}{3}.
\end{eqnarray} 
Electroneutrality condition is automatically satisfied in neutron stars. Spin concentrations just change accordingly: when some spin flips it increases the concentration of one type and decreases the concentration of opposite type.
Therefore, the energy of ``one-band'' SB~(\ref{eq-energy-smooth-dos}) based on a neutron star matter, which is stored in chemical potential spin splitting~$2 \eta$ is 
\begin{equation}
\delta E_n \approx \Omega D(\varepsilon_F) \eta^2 \approx \frac{3^\frac{2}{3} \Omega}{c \hbar}\Big(\frac{\rho}{\pi m_n}\Big)^\frac{2}{3}\eta^2 \ . 	  
\end{equation} 
Maximal energy which can be reversibly stored in ``spin battery'' based on such a neutron star corresponds to 100\% polarization of its neutrons, i.e.~$\eta = \varepsilon_F $. 
This allows us to estimate the maximal energy capacitance of one cubic millimeter of neutron star matter $\delta E_{n, max}(\Omega = 10^{-9} \text{m}^3) \sim 10^{42}$ Joules, which is $25$ orders of magnitude more than the energy of the most powerful thermonuclear bomb~\cite{ginness} tested by mankind~$2.4\cdot10^{17}$ Joules. 
The colossal energy reversibly stored for the whole neutron star with above parameters is~$\sim 10^{63}$~Joules.
Note, due to neutron has zero electric charge there is absent spin-orbit coupling and full spin polarization of the star does not lead to rotation of the star.

\section{Conclusion}
We have theoretically shown that spin batteries can surpass modern chemicals accumulators by orders of magnitude of energy capacitance due to high density of states of electrons in metals, especially in the case of metals with ``heavy fermions''. 
Fantastic scenario of using neutron star matter would allow reversibly store of colossal energy. 
We didn't find such estimations for neutron stars in literature. 


\section{Acknowledgement}
We thank L. A. Pastur for helpful discussions. P.~V.~P. acknowledge support of supported by Grant No. PGC2018-101355-B-I00 funded by MCIN/AEI/10.13039/501100011033 and by ``ERDF A way of making Europe''.

\bibliography{maximal_accumulation_eng_prb_6}

\end{document}